\documentclass[useAMS,usenatbib,usegraphicx]{mn2e}

\title[Pressure shifts and abundance gradients in DAZ white dwarf]
{Pressure shifts and abundance gradients in the atmosphere of the DAZ white dwarf GALEX~J193156.8+011745
\thanks{Based on observations made with ESO telescopes at the La Silla Paranal Observatory
under programme 283.D-5060.}}
\author[S. Vennes et al.]{S. Vennes$^{1}$\thanks{E-mail: vennes@sunstel.asu.cas.cz (SV); kawka@sunstel.asu. cas.cz (AK);
nemeth@sunstel.asu.cas.cz (NM)}, 
A. Kawka$^{1}$\footnotemark[2], and P. N\'emeth$^{1,2}$\footnotemark[2]\\
$^{1}$Astronomick\'y \'ustav AV \v{C}R, Fri\v{c}ova 298,CZ-251 65 Ond\v{r}ejov,
Czech Republic\\
$^{2}$Department of Physics and Space Sciences, 150 W. University Blvd, Florida Institute of Technology, Melbourne, FL 32901, USA
}

\begin{document}

\date{}

\pagerange{\pageref{firstpage}--\pageref{lastpage}} \pubyear{2010}

\maketitle

\label{firstpage}

\begin{abstract}
We present a detailed model atmosphere analysis of high-dispersion and
high signal-to-noise ratio spectra of the heavily polluted DAZ white dwarf
GALEX~J1931+0117. The spectra obtained with the VLT-Kueyen/UV-Visual Echelle Spectrograph
show several well-resolved Si\,{\sc ii} spectral lines enabling a 
study of pressure effects on line profiles. We observed large
Stark shifts in silicon lines in agreement with theoretical predictions and laboratory
measurements. 
Taking into account Stark shifts in the calculation of synthetic spectra we reduced the scatter
in individual line radial velocity measurements from $\sim 3$ to $\la 1$ km\,s$^{-1}$.
We present revised abundances of O, Mg, Si, Ca, and Fe based on a critical review of line broadening parameters
and oscillator strengths. The new measurements are generally in agreement with our
previous analysis with the exception of magnesium with a revised abundance a 
factor of two lower than previously estimated. The magnesium, silicon and iron abundances
exceed solar abundances, but the oxygen and calcium abundances are below solar.
Also, we compared the observed line profiles to synthetic spectra computed with variable accretion
rates and vertical abundance distributions assuming diffusive steady-state. 
The inferred accretion rates vary from $\dot{M}=2\times10^6$ for calcium to $2\times10^{9}$ g\,s$^{-1}$ for 
oxygen. We find that the accretion flow must be oxygen-rich while being deficient in calcium relative to
solar abundances.
The lack of radial velocity variations between two
measurement epochs suggests that GALEX~J1931+0117 is probably not in a close binary and that
the source of the accreted material resides in a debris disc.
\end{abstract}


\section{Introduction}

GALEX~J193156.8+011745 (GALEX~J1931+0117, thereafter) is a hydrogen-rich white dwarf 
discovered by \citet{ven2010a} and characterized by an opulent heavy-element line spectrum and an infrared excess.
The original low-resolution spectrum obtained with the 
New Technology Telescope (NTT) at La Silla Observatory showed a strong Mg\,{\sc ii}$\lambda$4481 doublet and weaker
silicon lines. Follow-up echelle spectroscopy obtained with the Very Large Telescope (VLT)-Kueyen
enabled a detailed abundance study. The near-solar abundances of oxygen, magnesium, silicon, calcium
and iron bear the signature of an external supply of material accreting onto the surface of the white dwarf. 
Based on available data, Vennes et al. concluded that the supply may originate from a close,
sub-stellar companion or from a cool debris disc. 

The presence of heavy elements in hydrogen-rich white dwarfs has variously been interpreted as
intrinsic to the white dwarf, or as extrinsic, i.e., supplied by the interstellar medium \citep{dup1993}, by a nearby companion as in
post-common envelope systems \citep{deb2006,kaw2008}, or by a debris disc \citep{zuc2003,kil2006,far2008}.
However, accretion from the  interstellar medium is unlikely because of supply shortages \citep{far2010a}.
In the extrinsic scenarios, the elements are accreted and diffused in the atmosphere and envelope of the star
\citep[see][]{fon1979,koe2009}. 
An intrinsic, or internal, reservoir of heavy elements is also possible, but in either scenario
a self-consistent solution of the diffusion equation must explore the effect
of radiative acceleration on trace elements \citep{cha1995a,cha1995b}.

As a class, the polluted DA white dwarfs, or DAZs, are often defined by the 
detection of the Ca\,{\sc ii} H\&K doublet in optical spectra of cooler objects \citep{zuc2003,koe2005}, or by
the detection of carbon and silicon in the
ultraviolet spectra of warmer objects \citep{dup2009a,dup2009b}. 
Exceptionally, Mg\,{\sc ii}$\lambda$4481 is, so far, only detected in a handful of warm white dwarfs
such as EG~102 \citep{hol1997}, GALEX~J1931+0117, and two warm white dwarfs from the Sloan Digital Sky Survey (SDSS)
that show evidence of dusty and gaseous discs \citep{gan2006,gan2007}. The presence of a
large concentration of magnesium in the last two objects helped establish a strong link between heavy element
pollution and dense circumstellar environments.
Moreover, an infrared excess that cannot otherwise be explained by a cool companion, may
be attributed to a dust ring as in the case of the white dwarf G29-38 \citep{gra1990}.

We present new high-dispersion spectroscopic observations that help elucidate the nature of the peculiar abundance
pattern in GALEX~J1931+0117 in support of the debris disc model.
Our observations are presented in Section 2, and a detailed line profile and abundance analysis
is presented in Section 3.  In Section 3.1 we compile and evaluate line broadening parameters and oscillator strengths.
In Section 3.2 we describe steady-state diffusion models and predicted abundance gradients in the stable atmospheres of hot
white dwarfs.
In Section 3.3 we present revised abundances based on a review of atomic data, 
we assess non-local thermodynamic equilibrium (non-LTE) effects on measured abundances, and
we constrain the accretion rates on the atmosphere of GALEX~J1931+0117.
Finally, we conclude in Section 4.

\section{Observations}

Following the identification of GALEX~J1931+0117 as a DAZ white dwarf we obtained a series of 
echelle spectra using standard settings with the UV-Visual Echelle Spectrograph (UVES) attached to the VLT-Kueyen.  
All spectra were obtained with the slit width set at $1\arcsec$, offering a resolving power $R\approx 46000$.
The exposure times were 1450~s.
The first series of spectra were obtained on UT 2009 Nov 12 (epoch 1, thereafter) with the dichroic \#2
and the ``HER\_5'' and ``BK7\_5'' filters on the blue and red arms, respectively.
The blue spectrum was centred at 4370\AA\ and covered the range 3757-4985\AA, and the
red spectra were centred at 7600\AA\ and covered the ranges 5698-7532 and 7660-9464\AA.
The first series of spectra were analysed by us \citep{ven2010a}.
The second series of spectra were obtained on UT 2010 Mar 15 (epoch 2, thereafter) with the dichroic \#1
and the ``HER\_5'' and ``SHP700'' filters on the blue and red arms, respectively.
The blue spectrum was centred at 3900\AA\ and covered the range 3290-4518\AA, and
the red spectra were centred at 5800\AA\ and covered the ranges 4788-5750 and
5839-6808\AA.

We remeasured the H$\alpha$ radial velocity by fitting Gaussian profiles to the
well-exposed, narrow line cores obtained at epochs 1 and 2, $v_{\rm H\alpha,1}=37.6\pm0.8$ km\,s$^{-1}$
and $v_{\rm H\alpha,2}=36.8\pm1.0$ km\,s$^{-1}$, respectively.
The two measurements are identical within errors.  
A lack of variations suggests that the DA white dwarf is not in a close binary system.
The average of the two measurements is $\bar{v}_{\rm H\alpha} = 37.2\pm0.6$ km\,s$^{-1}$.

Table~\ref{tbl-1} updates the spectral line identifications presented in
\citet{ven2010a}. A set of nine Fe\,{\sc ii} lines were observed  allowing us
to improve our earlier iron abundance measurement. Two strong Si\,{\sc ii} lines
are added to our list, but their line velocities deviate significantly from 
the rest velocity, set by H$\alpha$, by 6 and 7 km\,s$^{-1}$.
A closer examination of other Si\,{\sc ii} lines observed at epoch 1 also reveals similar radial
velocity shifts that shall be investigated in Section 3.3.1.
The sulfur lines S\,{\sc ii}$\lambda$5014.069, $\lambda$5032.447, and $\lambda$5432.815\AA\ were not detected.

\begin{table}
\begin{minipage}{85mm}
\centering
\caption{Additional UVES line identifications.}
\label{tbl-1}
\centering
\renewcommand{\footnoterule}{\vspace*{-15pt}}
\begin{tabular}{llcc}
\hline
 Ion & $\lambda$ \footnote{Continued from \citet{ven2010a}.}& E.W. & $v_{\rm bary}$   \\
      &    (\AA) & (m\AA) &  (km~s$^{-1}$) \\
\hline
 \mbox{Fe\,{\sc ii}} & 4233.172 & 7    &  36.9 \\
 \mbox{Fe\,{\sc ii}} & 4923.927 & 17   &  36.9 \\
 \mbox{Fe\,{\sc ii}} & 5001.959 & 22   &  39.5 \\
 \mbox{Fe\,{\sc ii}} & 5018.440 & 21   &  36.9 \\
 \mbox{Si\,{\sc ii}} & 5041.024 & 131  &  42.8 \\
 \mbox{Si\,{\sc ii}} & 5055.984\footnote{Blended with Si\,{\sc ii} $\lambda5056.317$.} & 233   & 44.5 \\
 \mbox{Fe\,{\sc ii}} & 5097.271 & 12   &  40.9 \\
 \mbox{Fe\,{\sc ii}} & 5100.727 & 27   &  38.8 \\
 \mbox{Fe\,{\sc ii}} & 5169.033 & 26   &  36.8 \\
 \mbox{Fe\,{\sc ii}} & 5227.481 & 17   &  42.4 \\
 \mbox{Fe\,{\sc ii}} & 5260.260 & 20   &  40.8 \\
\hline
\end{tabular} \\
\end{minipage}
\end{table}

\section{Analysis and discussion}

We adopted the model atmosphere analysis of \citet{ven2010a} that was based
on non-LTE calculations made with {\sc tlusty-synspec} \citep{hub1995,lan1995}. 
All models were computed at $T_{\rm eff}=20890$ K and $\log{g}=7.90$.
\citet{ven2010a} found the difference between
the best-fitting $(T_{\rm eff},\log{g})$ to the Balmer line series using
non-LTE heavy-element blanketed model atmospheres and the best-fitting
solution using pure-hydrogen LTE models to be within statistical errors ($\Delta T_{\rm eff}=
+120$ K, $\Delta \log{g}=-.04$).
Heavy elements contribute less than 0.1\%
of the total electron density and their combined opacities are relatively
modest in this range of effective temperatures compared to the strong Lyman and Balmer lines and continua.

The model atmosphere structures and synthetic spectra employed in the line profile analysis were computed assuming non-LTE, 
and with varying abundances of
oxygen, magnesium, silicon, calcium and iron. The spectral syntheses are computed with different 
atomic data sets described below (Section 3.1). 
Also, we computed spectral syntheses with depth-dependent trace element abundances that were 
prescribed by diffusion theory (Section 3.2).
Finally, we determine the extent of non-LTE effects in the atmosphere of GALEX~J1931$+$0117 by comparing our non-LTE analysis to LTE results (Section 3.3.2).

\subsection{Atomic data}

Our original spectral syntheses were computed with the line list available
on the CD-ROM No. 23 of \citet{kur1995}\footnote{CD23, accessed at {\tt http://www.cfa.harvard.edu/amp/ampdata\\/kurucz23/sekur.html}.}. For each ion, we now compare the oscillator strengths ($f_{\rm ij}$) 
and Stark line broadening parameters ($\Gamma$) listed by \citet{kur1995} to the best available
theoretical and experimental data. Data on line oscillator strengths are also available at
the National Institute of Standards and Technology (NIST).\footnote{Accessed at {\tt http://www.nist.gov/physlab/data/asd.cfm}.}

We noted that the isotopic shift in the $^{26-24}$Mg\,{\sc ii}$\lambda$2798 doublet is only $+0.85$ km\,s$^{-1}$ \citep{dru1980}
and, therefore, we do not expect observable effects on the Mg\,{\sc ii}$\lambda$4481 doublet in 6\,km\,s$^{-1}$-resolution 
spectra. Similarly, other isotopic shifts (e.g., $^{30}{\rm Si}/^{28}{\rm Si}$) may be neglected 
in the present study \citep[see][]{ber2003}.

\subsubsection{Si\,{\sc ii}}

The rich silicon line spectrum in GALEX~J1931+0117 prompted a detailed review of
available data. Table~\ref{tbl-2} lists and compares $f_{\rm ij}$ from popular
data compilations (CD23 and NIST) to homogeneous theoretical or experimental data sets.
We noted considerable variations in Si\,{\sc ii} oscillator strengths, in particular
in the $\lambda$3862 triplet and in $\lambda$5041.024. Discrepancies of
the order of 40 to 50\% would affect individual abundance measurements in equal measures.
For example, the ratio of CD23 data to experimental $f_{\rm ij}$ values is 1.06 but varies with a standard
deviation $\sigma=$38\% while the NIST data that are largely based on the experimental values
vary by only 8\% with an average ratio of 0.98, and the theoretical data \citep{art1981} vary by 23\% with an
average ratio of 0.98. The adoption of one data set over another
will not have a large effect on the average abundance, but individual line measurements are
less reliable.

\begin{table}
\centering
\begin{minipage}{80mm}
\caption{Si\,{\sc ii} oscillator strengths ($f_{\rm ij}$).}
\label{tbl-2}
\centering
\renewcommand{\footnoterule}{\vspace*{-15pt}}
\begin{tabular}{cllll}
\hline
$\lambda_0$ & CD23\footnote{Compiled by \citet{kur1995}.} & NIST & Theory\footnote{\citet{art1981}.} & Exp.\footnote{\citet{mat2001}.}\\
(\AA)      &         &         &           &    \\
\hline
3853.665   & 0.0076  & 0.0114  &  0.0090   & 0.0131$\pm$0.0014   \\
3856.018   & 0.0462  & 0.0654  &  0.0535   & 0.073$\pm$0.011   \\
3862.595   & 0.0381  & 0.0437  &  0.0445   & 0.039$\pm$0.007   \\
4128.054   & 0.518   & 0.571   &  0.600    & 0.57$\pm$0.18   \\
4130.872   & 0.025   & 0.0275  &  0.0285   & ...   \\
4130.894   & 0.499   & 0.594   &  0.570    & 0.6$\pm$0.2   \\
5041.024   & 0.977   & 0.534   &  0.765    & 0.53$\pm$0.09   \\
5055.984   & 0.979   & 0.834   &  0.690    & 0.83$\pm$0.16   \\
5056.317   & 0.109   & 0.0805  &  0.0768   & ...   \\
5957.559   & 0.250   & 0.298   &  0.233    & ...  \\
5978.930   & 0.252   & 0.303   &  0.233    & ...  \\
6347.109   & 0.991   & 0.705   &  0.745    & 0.66$\pm$0.11  \\
6371.371   & 0.497   & 0.414   &  0.373    & 0.46$\pm$0.08   \\
\hline
\end{tabular} \\
\end{minipage}
\end{table}

Strong saturated lines are sensitive to line broadening parameters. Table~\ref{tbl-3} lists
full-width at half-maximum (FWHM$\equiv 2w$) Stark widths and shifts due to electron impacts for strong Si\,{\sc ii} optical lines.
\citet{kur1995} tabulate the circular frequency per electron $\Gamma$ that we converted into the FWHM at
$n_e=10^{17}$ cm$^{-3}$ using the formula:
\begin{equation}   
2w = \frac{1}{2\pi}\,\Gamma\, \frac{\lambda^2}{c}\,n_e,
\end{equation}   
where $c$ is the speed of light. \citet{lan1988} tabulates FWHM values for electron and proton
impacts separately. For most lines the proton contribution to the total width 
is $\approx$10\% up to 15\%. The FWHM from \citet{kur1995} are on average
43\% larger than the experimental values with a standard deviation of 44\% while
the values from \citet{lan1988} are only 6\% lower than the experimental values
with a standard deviation of 21\%. 

The excellent agreement between \citet{lan1988} and the experiments prompted us to explore
two options in
computing detailed silicon line spectra. In option 1 we adopted the 
silicon oscillator strengths from \cite{art1981} and the line broadening
parameters for electrons and protons from \citet{lan1988}. The line broadening parameters are tabulated at
5, 10, 20, and 40$\times10^3$ K with estimated uncertainties of less than 20\% at
$n_e=10^{17}$ cm$^{-3}$ (or a depth $\tau_R\approx 2$ in the atmosphere). Also, the effect of Stark shifts are included
using the experimental data of \citet{gon2002} although we have 
no information on the scaling of $d_e$ with temperature
or on the magnitude of Stark shifts due to ions (protons in this case).
The effect of Stark shifts is apparent in radial velocity measurements of the Si\,{\sc ii}
$\lambda\lambda$5041-5055 and $\lambda\lambda$5967-5978 multiplets \citep[see Table~\ref{tbl-1} and][]{ven2010a}.
In option 2, we employed the data ($f_{\rm ij}$ and $\Gamma$) compiled by \citet{kur1995},
and neglected the effect of Stark shifts.

\begin{table}
\centering
\begin{minipage}{85mm}
\caption{Si\,{\sc ii} Stark widths (FWHM) and shifts.}
\label{tbl-3}
\centering
\renewcommand{\footnoterule}{\vspace*{-15pt}}
\begin{tabular}{ccccccc}
\hline
$\lambda_0$ & CD23\footnote{Compiled by \citet{kur1995}.} & &  Theory\footnote{\citet{lan1988} calculated at $n_e=10^{17}$ cm$^{-3}$ and $T=20\times10^3$ K.} & & \multicolumn{2}{c}{Experiments \footnote{At $n_e=10^{17}$ cm$^{-3}$ and $T=16-20\times10^3$ K \citep{gon2002,les2009}.}} \\
\cline{2-2} \cline{4-4} \cline{6-7}
           & $2w_e$ & &   $2w_e$     & &   $2w_e$         &  $d_e$       \\
(\AA)      & (\AA) & &   (\AA)      & &   (\AA)          &   (\AA)      \\
\hline
3853.665   & 0.97 & &   0.42       & &   0.52$\pm$0.17  &  $-$0.17$\pm$0.05   \\
3856.018   & 0.97 & &   0.42       & &   0.50$\pm$0.11  &  $-$0.05$\pm$0.01   \\
3862.595   & 0.97 & &   0.42       & &   0.50$\pm$0.13  &  $-$0.03$\pm$0.01   \\
4128.054   & 1.22 & &   0.96       & &   0.97$\pm$0.15  &  $-$0.23$\pm$0.03   \\
4130.894   & 1.22 & &   0.96       & &  1.01$\pm$0.11   &  $-$0.20$\pm$0.02   \\
5041.024   & 2.69 & &   1.95       & &  2.54$\pm$0.18   &  +0.86$\pm$0.06      \\
5055.984   & 2.71 & &   1.95       & &  2.58$\pm$0.34   &  +0.93$\pm$0.12      \\
5957.559   & 2.32 & &   2.27       & &  2.72$\pm$0.54   &  +1.32$\pm$0.26 \\
5978.930   & 2.33 & &   2.27       & &  2.78$\pm$0.42   &  +1.19$\pm$0.18 \\
6347.109   & 1.95 & &   1.44       & &  1.13$\pm$0.26   &  $-$0.31$\pm$0.07   \\
6371.371   & 1.97 & &   1.44       & &  0.99$\pm$0.25   &  $-$0.29$\pm$0.07   \\
\hline
\end{tabular} \\
\end{minipage}
\end{table}

\subsubsection{Mg\,{\sc ii}}

Unlike silicon, the magnesium abundance measurement is based on a single
doublet. The red Mg\,{\sc ii}$\lambda$7989 multiplet shows evidence of Stark shifts \citep{ven2010a}
although the spectrum is particularly noisy in the vicinity of the multiplet.
On the other hand, the strong Mg\,{\sc ii}$\lambda$4481 doublet is well exposed. Because of
its strength, the line profile is particularly sensitive to broadening parameters,
although quoted oscillator strength values are consistent \citep[NIST and][]{kur1995}
within $\la2$\%. 

The compilation of \citet{kur1995} does not provide a Stark width for the Mg\,{\sc ii}$\lambda$4481 doublet
which is then estimated in {\sc synspec} using the formula \citep{cas2005}
\begin{equation}   
\Gamma=10^{-8}\,n_{\rm eff}^5,
\end{equation}   
where $\Gamma$ is expressed in rad\,s$^{-1}$\,cm$^{-3}$ and $n_{\rm eff}$ is the effective quantum number in the hydrogenic approximation 
\begin{equation}   
n^2_{\rm eff}= (Z+1)^2 \frac{13.595}{E_i},
\end{equation}   
where $E_i$ is the ionization energy (eV) of the upper level of the transition and $Z$ is the ionic charge of the trace element.
The $\Gamma$ value is then converted into the FWHM $2w$.
Table~\ref{tbl-4} shows that theoretical and experimental values of the widths and shifts due to electrons
are in agreement but that the estimated width using the classical formula is in error. 
The Stark shift due to electrons compares well with experiments \citep[see][]{les2009}, although
the shift due to protons is predicted to dominate \citep{dim1995}.
Incidentally, \citet{dim1995} also predict a pressure shift in the Mg\,{\sc ii}$\lambda$7989 multiplet
of opposite sign and a factor of $\approx$8 larger than in the Mg\,{\sc ii}$\lambda$4481 doublet,
in qualitative agreement with observations reported here and by \citet{ven2010a}.

In step with our investigation of silicon lines, we also developed two options for the analysis of the
Mg\,{\sc ii}$\lambda$4481 doublet. In option 1 we adopted the NIST oscillator strengths and
the line broadening and shift parameters (proton and electron) from \citet{dim1995}. 
In option 2, we adopted the oscillator strengths from \citet{kur1995} and the classical line
broadening parameter while neglecting pressure shifts.

\begin{table}
\centering
\begin{minipage}{85mm}
\caption{Mg\,{\sc ii} Stark widths (FWHM) and shifts.}
\label{tbl-4}
\centering
\renewcommand{\footnoterule}{\vspace*{-15pt}}
\begin{tabular}{cccccccc}
\hline
$\lambda_0$ & $\Gamma$ Est.\footnote{Estimated using the classical formula and $n_{\rm eff}=4$.} & &  \multicolumn{2}{c}{Theory\footnote{\citet{dim1995} calculated at $n_e=1.35\times10^{17}$ cm$^{-3}$ and $T=16.9\times10^3$ K.}} & & \multicolumn{2}{c}{Experiments \footnote{Measured at $n_e=1.35\times10^{17}$ cm$^{-3}$ and $T=16.9\times10^3$ K \citep{dje2005,les2009}.}} \\
\cline{2-2} \cline{4-5}  \cline{7-8}
         &  $2w_e$ & &    $2w_e$ & $d_e$    & &   $2w_e$         &  $d_e$       \\
(\AA)    &   (\AA) & &    (\AA)  & (\AA)    & &   (\AA)          &   (\AA)      \\
\hline
4481.126   & 1.09 & &    2.50   & $-0.0883$ & &   2.95          &  $-0.085$    \\
\hline
\end{tabular} \\
\end{minipage}
\end{table}

\subsubsection{Ca\,{\sc ii}}

Table~\ref{tbl-5} compares theoretical and experimental results for the Ca\,{\sc ii}$\lambda$3933 
resonance line. Stark widths from various sources including \citet{kur1995} are in agreement 
although the pressure shifts calculated by \citet{dim1992} are a factor of 3 smaller
than measured \citep{sre1993}. Because the measured Ca\,{\sc ii}$\lambda$3933 shift is
relatively small \citep[see][]{ven2010a} it may not be measurable in our spectra and the lack of
accuracy in these parameters would not affect the abundance and radial velocity analysis.

\begin{table}
\centering
\begin{minipage}{85mm}
\caption{Ca\,{\sc ii} Stark widths (FWHM) and shifts.}
\label{tbl-5}
\centering
\renewcommand{\footnoterule}{\vspace*{-15pt}}
\begin{tabular}{cccccccc}
\hline
$\lambda_0$ & CD23\footnote{Compiled by \citet{kur1995} and accessed at {\tt http:\\//www.cfa.harvard.edu/amp/ampdata/kurucz23/sekur.html}.} & &  \multicolumn{2}{c}{Theory\footnote{\citet{dim1992} calculated at $n_e=1.76\times10^{17}$ cm$^{-3}$ and $T=43\times10^3$ K.}} & & \multicolumn{2}{c}{Experiments \footnote{At $n_e=1.76\times10^{17}$ cm$^{-3}$ and $T=43.0\times10^3$ K \citep{sre1993}.}} \\
\cline{2-2} \cline{4-5}  \cline{7-8}
         &  $2w_e$ & &    $2w_e$ & $d_e$    & &   $2w_e$         &  $d_e$       \\
(\AA)    &   (\AA) & &    (\AA)  & (\AA)    & &   (\AA)          &   (\AA)      \\
\hline
3933.663   & 0.248 & &    0.300  & $-0.048$ & &  0.286          &  $-0.14$      \\
\hline
\end{tabular} \\
\end{minipage}
\end{table}

As part of option 1 we adopt the theoretical broadening and shift parameters of \citet{dim1992}
and in option 2 we adopt again the compilation of \citet{kur1995} without pressure shifts.

\subsubsection{O\,{\sc i} and Fe\,{\sc ii}}

The analysis of weak spectral lines is sensitive to the assumed oscillator strengths
but it is less sensitive to broadening parameters, and we will not distinguish the two
options.

The O\,{\sc i}$\lambda$7773 oscillator strengths tabulated by \citet{kur1995} and NIST
differ by $\approx$10\% which implies a minimum uncertainty in oxygen abundance measurements
of the same order.
\citet{ben1996} compared calculated O\,{\sc i} Stark width to \citet{gri1974} and
to experiments. They found excellent agreement between their calculations and the
experimental values although they do not provide data for the  O\,{\sc i}$\lambda$7773
multiplet. The Stark width tabulated by \citet{gri1974} and \citet{kur1995}
agree to within a few percent at $T=20000$ appropriate for our analysis.
We adopted the oscillator strengths and Stark widths tabulated by \citet{kur1995}.

Five of the Fe\,{\sc ii} lines detected in GALEX~J1931+0117 are listed at NIST
while atomic data for all lines are listed by \citet{kur1995}. 
In this case,
we adopted the oscillator strengths and Stark widths tabulated by \citet{kur1995}.

\subsection{Diffusion in white dwarf atmospheres}

A tenet of the accretion-diffusion scenario is that photospheric abundances 
reach a steady-state level in the line-forming region. The abundance gradient is assumed
to be zero in convective layers, but it is a function of the diffusion time scale in radiative layers
as in the case of GALEX~J1931+0117. 

Assuming accretion steady-state the mass-continuity equation may be written as 
\begin{equation}   
c_2 \equiv \frac{n_2}{n_1} = \dot{M_2} \frac {1}{4\pi R^2}\frac{1}{A_2}\frac{\tau_{1,2}}{m},
\end{equation}   
where $c_2$ is the abundance of element ``2'' relative to the main constituent ``1'',
$\dot{M_2}$ is the mass accretion rate of element ``2'' (g\,s$^{-1}$), $R$ is the stellar radius (cm), $A_2$ is the
atomic weight of element ``2'' (a.m.u), 
\begin{equation}
\tau_{1,2} = m/(\rho v_{1,2}) 
\end{equation}
is the diffusion time scale of element ``2'' (s), $m$ and $\rho$ are the mass loading (g\,cm$^{-2}$) and density 
(g\,cm$^{-3}$) in the
atmosphere, and $v_{1,2}$ is the diffusion velocity (cm\,s$^{-1}$)
given by the diffusion equation.
Assuming that element ``1'' is fully ionized hydrogen ($A_1=1,\ Z_1=1$) and that $c_2<<1$, then the diffusion equation \citep[see][]{ven1988} simplifies to
\begin{equation}   
v_{1,2} = D_{1,2}\Big{(} -\frac{1}{c_2}\frac{d\,c_2}{d\,r} + (1-A_2) \frac{m_p\,g}{kT}+ A_2 \frac{m_p\,g_r}{kT}\Big{)},
\end{equation}   
where $D_{1,2}$ is the diffusion coefficient of trace element ``2'' 
relative to the main constituent \citep[see][]{ven2010b}.

The first term on the right-hand side is the actual abundance gradient. We show in Section 3.3.3 that $c_2$ is a slowly
varying function of density, $c_2\propto\rho^\alpha$ and $|\alpha|\la 0.5$, therefore
\begin{equation}   
\Big{|}\frac{1}{c_2}\frac{d\,c_2}{d\,r} \Big{|} \la -\frac{1}{\rho}\frac{d\,\rho}{d\,r} = \frac{d\,\rho}{d\,m},
\end{equation}   
where $dm=-\rho\,dr$.
The second term on the right-hand side of the diffusion equation is proportional to the surface gravity $g$
and the last term is 
proportional to the radiative acceleration $g_r$.
Employing the run of density $\rho$ and temperature $T$ as a function of mass loading $m$ in a model
atmosphere appropriate for GALEX~J1931+0117, it can be shown that
gravitational settling acts unimpeded as long as $c_2<<1$, i.e., that 
\begin{equation}   
\Big{|} \frac{d\,\rho}{d\,m} \Big{|} << \Big{|}(1-A_2)\,\frac{m_p\,g}{kT}\Big{|}.
\end{equation}   
The abundance gradient term is $\la 2$\% of the gravity
term and does not effectively counteract the gravitational pull. Similarly, radiative forces \citep{cha1995a,cha1995b} are
relatively weak at $T_{\rm eff}\approx 20\times10^3$K and cannot raise the  abundances to levels
observed in GALEX~J1931+0117. The silicon abundance potentially supported by radiation is at least 3.5 orders of magnitude below
the observed abundance, and the radiative forces on oxygen, magnesium, calcium, and iron are negligible. 

In summary, unless a diffusion steady-state cannot be reached, the vertical abundance distribution in the atmosphere
of an accreting white dwarf follows a relation of the form $c_2 \propto \tau_{1,2}/m$ and may depart significantly
from a homogeneous distribution.
Longitudinal homogeneity is assumed although abundance spots may in principle occur.

\subsection{Line profile analysis}

Detailed line profiles are computed and convolved to the instrumental resolution ($R\approx 46000$).
The separate routines were introduced into the spectral synthesis code to treat Stark shifts
self-consistently. The Voigt profiles, that already include a treatment of Doppler width and
damping functions ($\Gamma_{\rm nat}+\Gamma_{\rm Stark}$) have been modified to include
the effect of Stark shifts. The line centre is shifted linearly with electronic and proton density
and proportionally to the Stark shift $d$ tabulated at a reference density.
Table~\ref{tbl-6} lists the revised abundances.

\subsubsection{Line broadening and pressure shift}

\begin{figure}
\includegraphics[width=0.99\columnwidth]{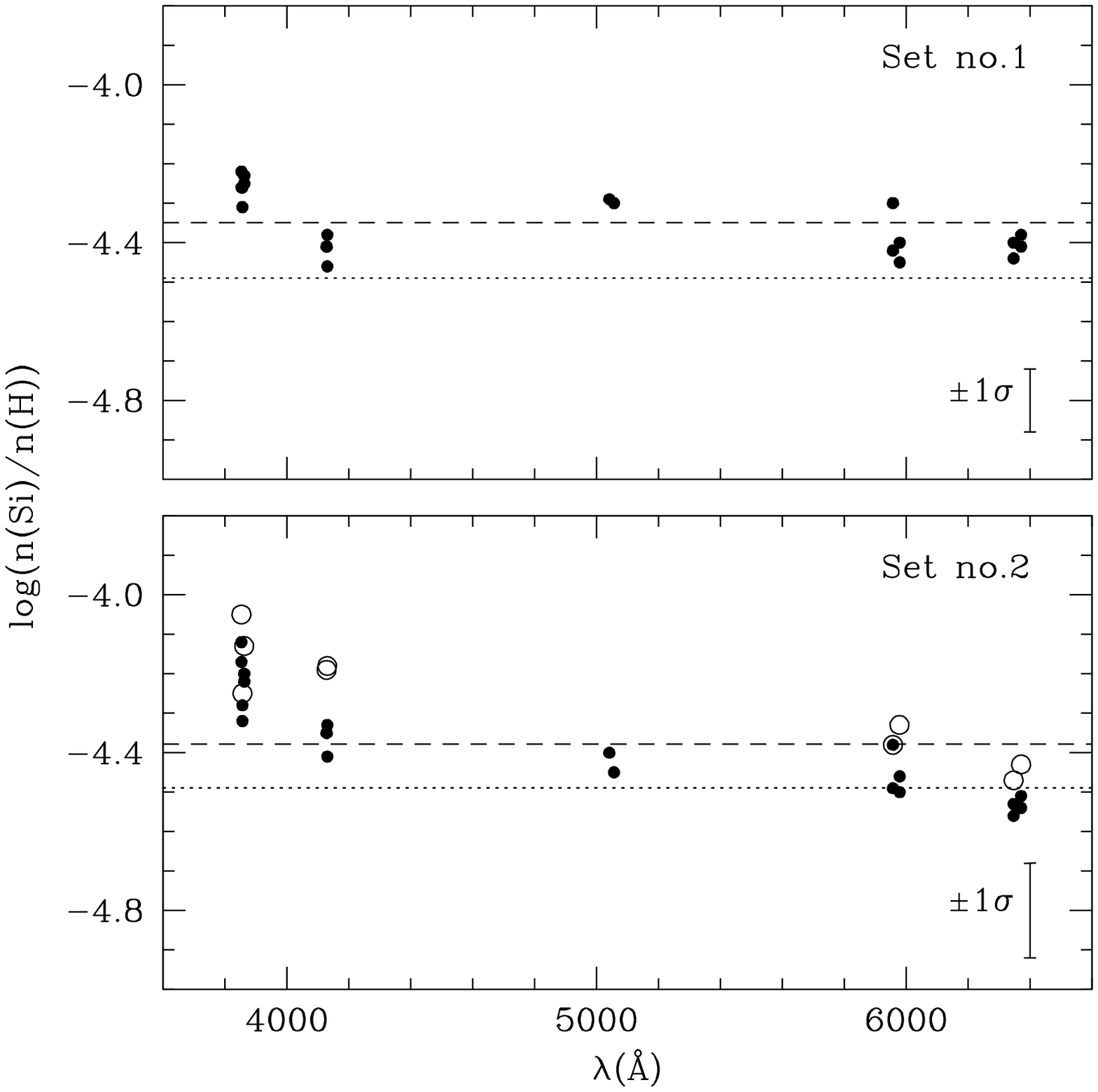}
\caption{Abundances measured in LTE with individual lines and using two separate atomic
data sets. Set number 1 employs line broadening calculations of \citet{lan1988} and
oscillator strengths from \citet{art1981} and includes the effect of Stark shifts.
Set number 2 employs data compiled by \citet{kur1995} but excludes Stark Shifts. 
Twenty abundance measurements (full circles) with set number 1
are shown in the upper panel, while the same abundance measurements using set number 2 are shown in the
lower panel along with the non-LTE measurements (open circles) collected by \citet{ven2010a} and showing
the extent of non-LTE effects on the silicon abundance measurements. Solar abundance
($\log{n({\rm Si})/n({\rm H})_\odot}=-4.49$) is shown with dotted lines
\citep{asp2005} and the measured averages are shown with dashed lines.
Standard deviations ($\pm1\sigma$) for each data set are shown in the
lower right corner (see text).
}\label{fig1}
\end{figure}

\begin{figure}
\includegraphics[width=0.99\columnwidth]{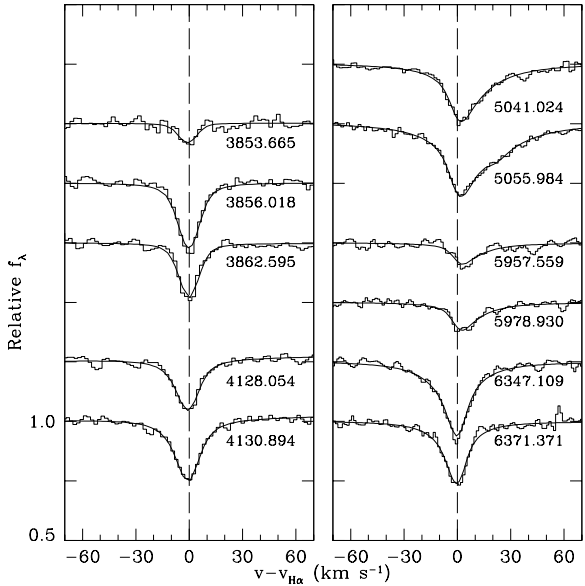}
\caption{Line profile fits of spectra obtained at epoch 2 and corresponding to abundances depicted in the top panel of Figure~\ref{fig1}. 
}\label{fig2}
\end{figure}

In the present analysis we assume both longitudinal and vertical abundance uniformity.
The lack of abundance variations between two observation epochs favours a uniform spread of trace
elements across the surface although additional observations are required to rule out 
a mere coincidence in rotational phases. Figure~\ref{fig1} shows abundance measurements
obtained for nine Si\,{\sc ii} spectral lines in 20 separate measurements and Figure~\ref{fig2}
shows sample line profile fits. 

The revised silicon abundance is $-0.05$ dex (or 10\%) lower than estimated by \citet{ven2010a} 
mainly because of the additional lines included in the analysis.
In fact, the abundance measurements obtained assuming option 1 were on average 7\% larger than
with option 2. However, as proposed in option 1,
adopting the silicon oscillator strengths and line-broadening parameters from \citet{lan1988} and \citet{art1981}, respectively, 
reduced the scatter in abundance measurements based on 20 separate measurements.
The scatter in abundance measurements obtained with option 1 was
$\sigma = 0.08$ while the scatter obtained with option 2
was larger, $\sigma=0.13$, while the measurements showed a definite wavelength-dependent
trend (Fig.~\ref{fig1}). 

\begin{figure}
\includegraphics[width=0.99\columnwidth]{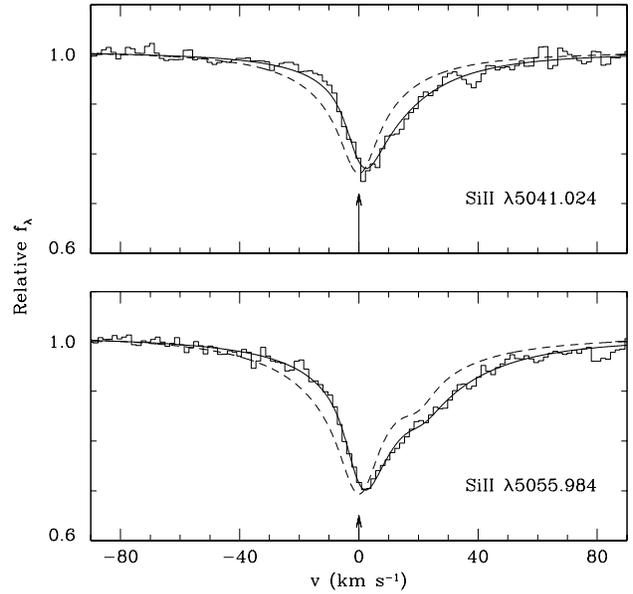}
\caption{Spectra obtained at epoch 2 are fitted with model line profiles including Stark shifts (full lines) or excluding
the effect (dashed lines). The velocity shift
in the observed line profiles is well reproduced by models including Stark shifts.
}\label{fig3}
\end{figure}

Moreover, the
inclusion of Stark shifts in line profile modelling reduced the scatter in Si\,{\sc ii} line velocity measurements.
The standard deviation of the measurements is $\sigma_v = 2.8$ without shifts down to $\sigma_v = 0.9$ km\,s$^{-1}$
with Stark shifts included. The average line velocity is $\bar{v} = -0.8$
relative to $v_{\rm H\alpha}= 37.2$ km\,s$^{-1}$ without Stark shifts and $0.0$ 
km\,s$^{-1}$ with Stark shifts included.  
The modelling of the spectral lines with the largest expected Stark shifts has been considerably
improved by taking this effect into account (Fig.~\ref{fig2}). Figure~\ref{fig3} compares 
Si\,{\sc ii} line profiles computed with and without Stark shifts to the observed spectra. 
The zero velocity is set by the H$\alpha$ line core ($+37.2$ km\,s$^{-1}$).
The line asymmetry in $\lambda$5055 is due in part to the presence of a weaker Si\,{\sc ii} line over the extended red wing,
but the asymmetric profile in $\lambda$5041 is entirely due to the Stark shift.
The velocity discrepancy notable for these two lines is entirely resolved using Stark shifted
profiles.

The calcium abundance was revised slightly down by $-0.06$ dex (or 14\%) mainly because of a small increase
in the adopted Stark width. The effect of pressure shift was not noticeable. We 
estimated the error on the abundance measurements by assuming an uncertainty of 25\%
on the broadening parameter. The corresponding uncertainty on the abundance is 0.05 dex or 10\%. 

Our revised magnesium abundance is significantly lower
than originally estimated by \citet{ven2010a} entirely because of an upward revision of
the Stark width by a factor of two in better agreement with experimental measurements
(Table~\ref{tbl-4}). We also estimated the abundance error assuming
a 25\% uncertainty, i.e., at least the difference between theory and experiments, on the broadening parameter.
The corresponding uncertainty on the abundance is 0.07 dex or 15\%.

The iron abundance, now based on nine separate line measurements, is, within errors, identical to
the original estimate. The error was set to the standard deviation of the set of measurements.
The procedure
adequately propagates uncertainties in broadening parameters and oscillator strengths. However,
variations of 25\% on the Stark width had no effect on the line equivalent widths. Similarly,
the oxygen abundance is identical to the original estimate, and the error was set to the
corresponding uncertainty in oscillator strengths (10\%). Variations in the Stark width had no
effect on the equivalent widths because weak lines such as Fe\,{\sc ii} and O\,{\sc i} lie on
the linear part of the curve-of-growth and are insensitive to broadening parameters unlike
strong Mg\,{\sc ii}, Si\,{\sc ii}, or Ca\,{\sc ii} lines.

\subsubsection{Non-LTE effects on abundance measurements}

We measured the abundances using the grid of LTE spectra and compared them to the non-LTE results.
We found that 
non-LTE effects are the largest in highly-excited Si\,{\sc ii} lines. The average non-LTE shift on abundance measurements is 
\begin{displaymath}
\Delta \log{\rm Si/H} = \log{\rm Si/H}_{\rm NLTE}-\log{\rm Si/H}_{\rm LTE} = +0.08\pm0.03
\end{displaymath}
The abundance shift is $\approx +0.05$ for spectral lines with the lower levels 3s3p$^2$($^2$D) or 3s$^2$4s($^2$S), 
and $\approx +0.11$ for highly-excited lines with the lower levels 3s$^2$3d($^2$D) or 3s$^2$4p($^2$P$^{\rm o}$). 

A similar comparison of magnesium, calcium, and iron abundances show that non-LTE effects are small:
\begin{displaymath}
\Delta \log{\rm Mg/H} = -0.02
\end{displaymath}
\begin{displaymath}
\Delta \log{\rm Ca/H} = -0.02
\end{displaymath}
\begin{displaymath}
\Delta \log{\rm Fe/H} = +0.01
\end{displaymath}
Finally, the abundance of oxygen measured with highly-excited multiplet O\,{\sc i}$\lambda7773$
is relatively large:
\begin{displaymath}
\Delta \log{\rm O/H} = -0.13
\end{displaymath}
In summary, non-LTE effects are notable in highly-excited lines of non-dominant species such
as O\,{\sc i}, but are negligible in strong lines of dominant species such
as Ca\,{\sc ii} or Mg\,{\sc ii}. Highly-excited levels of Si\,{\sc ii} are also affected.

\subsubsection{Effect of abundance gradient on line profiles}

We investigated the effect of vertical abundance inhomogeneities in the calculation of a
spectral line syntheses. The depth-dependent abundance was calculated by fixing the accretion rate
$\dot{M}$ and the stellar radius $R$ and varying the diffusion time-scale $\tau_{1,2}$ as a function of mass
loading $m$ as described in Section 3.2. We computed the time-scale and corresponding 
diffusion velocity following \citet{ven2010a}. The effect of neutral species on element-averaged
velocities was investigated by excluding them from the diffusion velocity calculations and comparing
the resulting abundance profiles.

\begin{figure*}
\includegraphics[width=1.00\columnwidth]{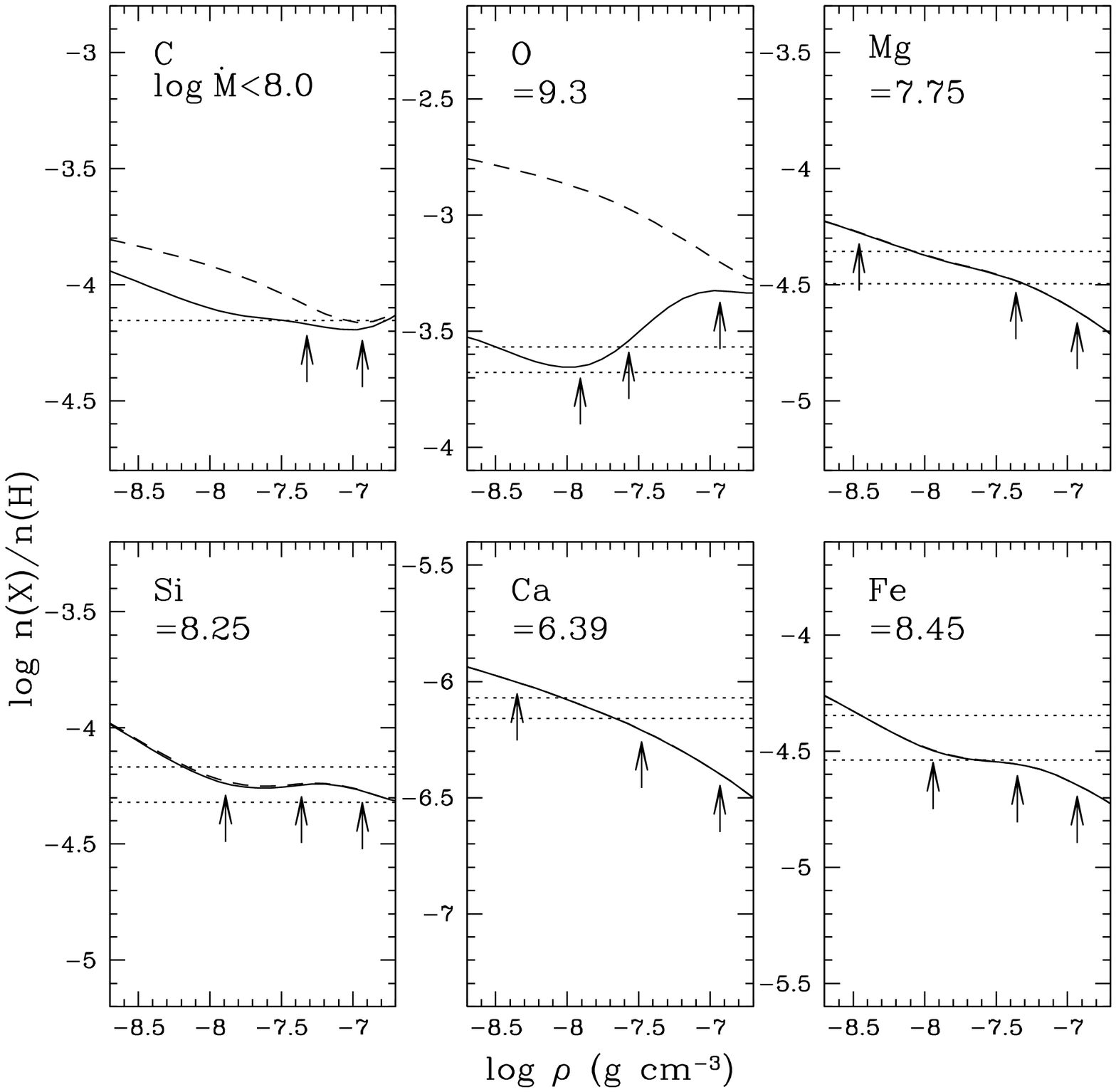}
\includegraphics[width=1.00\columnwidth]{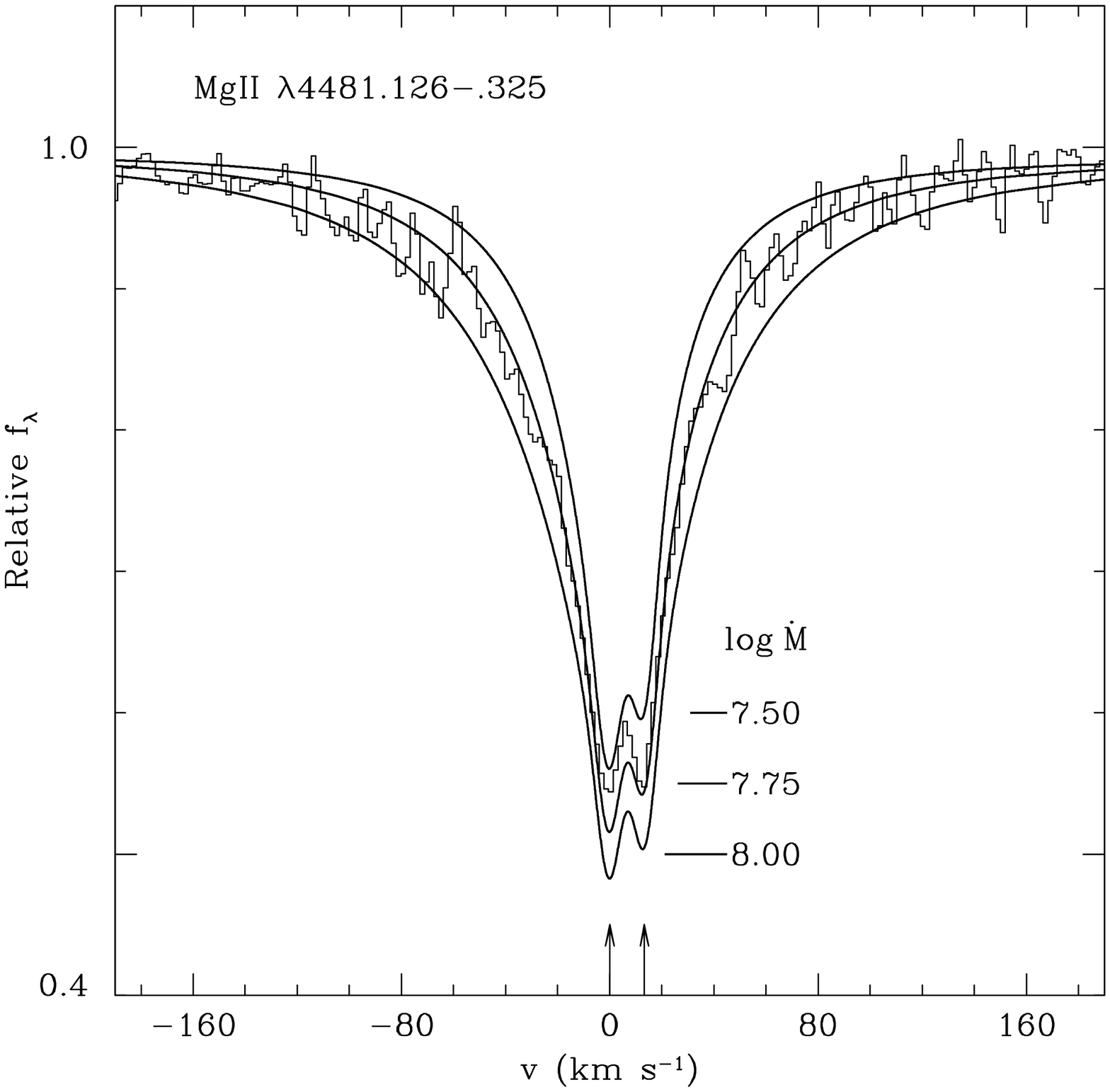}
\caption{(Left) vertical abundance distribution of elements in the atmosphere of GALEX~J1931+0117 computed assuming
diffusive steady-state (Section 3.2) and typical accretion rates $\dot{M}$ in g\,s$^{-1}$ (full lines). The effect of neutral
species on element-averaged diffusion velocities and vertical abundance distributions is shown by excluding neutral species from the
calculations (dashed lines). The profiles are compared
to measured abundances from Table~\ref{tbl-6} (dotted lines). The locations of 
optical depth $2/3$ at
$\lambda=1260$\AA\ (rightmost) and in the line wings and center (leftmost) are shown with arrows. (Right)
spectrum obtained at epoch 2 (jagged line) compared to model line profiles assuming Mg accretion rates 
$\dot{M}=3.2$, 5.6, and $10\times10^7$\,g\,s$^{-1}$.
}\label{fig4}
\end{figure*}

Figure~\ref{fig4} (left) shows that in the line forming region the abundance ratio $c_2\equiv n_2/n_1$ can be
approximated by $c_2 \propto \rho^{\alpha}$ where $-0.44\la \alpha \la 0.54$, or $|\alpha|\la 0.5$. The simulations
confirm that, in this context, the abundance gradients are not steep enough to counteract gravity (Section 3.2).
The effect of neutral species on element-averaged diffusion velocities are more marked in the cases of carbon
and oxygen because of the relatively high C\,{\sc i} and O\,{\sc i} ionization potentials. In the case of oxygen the
assumed location of the photosphere bears heavily of the derived accretion rate and other elements may be
affected as well. In order to remedy the situation we 
computed new emerging line profiles parametrized by a constant accretion rate, i.e., an accretion steady-state, 
rather than a constant abundance throughout the photosphere. 
These non-LTE spectral syntheses were calculated by adopting the structure ($T,\rho$)
of the best-fitting {\it homogeneous} model and introducing the abundance
distribution prescribed by Equation 4.
The observed spectra were then fitted with the model grid and
the accretion rates were measured explicitly.

Table~\ref{tbl-7} lists measured accretion rates for C, O, Mg, Si, Ca, and Fe. 
The quoted errors are statistical only and do not include systematic errors
in modelling the diffusion flow. Calculated concentrations in the line
forming region are inversely proportional to diffusion coefficients and errors
would scale accordingly, i.e., an error of 50\% on theoretical diffusion
coefficients would propagate to calculated concentrations.
Figure~\ref{fig4} (left) shows the corresponding abundance profiles. 
The inferred abundances at three different locations in the photospheres (marked
with arrows) are
in good agreement with the abundance derived from homogeneous models in the case silicon, and they deviate slightly
in the case of magnesium and iron. Figure~\ref{fig4} (right) shows model spectra compared to the observed spectrum at
three accretion rates. Although abundance inhomogeneities may affect line profiles, in particular in the relative
strength of line core and line wings, a comparison of ultraviolet to optical line strengths may be more revealing,
particularly for ultraviolet oxygen lines.

The inferred chemical composition of the accretion flow may be estimated by calculating the abundances relative to
silicon
\begin{equation}   
\frac{n{\rm (X)}}{n{\rm (Si)}}\Big{|}_{\rm flow} = \frac{\dot{M}{\rm (X)}}{A{\rm (X)}}\Big{(}\frac{\dot{M}{\rm (Si)}}{A{\rm (Si)}}\Big{)}^{-1},
\end{equation}   
where X refers to C, O, Mg, Ca, and Fe. Table~\ref{tbl-7} list the corresponding abundance ratios relative to solar
ratios. The results show that oxygen must be supplied in larger proportions that afforded by a solar composition, and
that silicon is also supplied in larger proportions than carbon, magnesium, or calcium. Iron is apparently supplied
in solar proportions relative to oxygen.

In summary, a line profile analysis parametrized by accretion rates rather than constant abundances reveals
an oxygen enrichment in the accretion flow. Diffusion calculations show the dominant effect of neutral species on 
abundance profiles. 

\begin{table}
\centering
\begin{minipage}{85mm}
\caption{Revised non-LTE abundances in GALEX~J1931+0117.}
\label{tbl-6}
\centering
\renewcommand{\footnoterule}{\vspace*{-15pt}}
\begin{tabular}{ccc}
\hline
X & $n({\rm X})/n({\rm H})$  & [X/H] \footnote{[X/H]$\equiv\log{n({\rm X})/n({\rm H})}-\log{n({\rm X})/n({\rm H})}_\odot$} \\
\hline
C\footnote{From \citet{ven2010a}.}     & $<7\times10^{-5}$ & $<-0.54$ \\
O     & $2.4\pm0.3\times10^{-4}$  & $-0.28$ \\
Mg    & $3.8\pm0.6\times10^{-5}$  & $\phantom{+}0.05$ \\
Si    & $5.8\pm1.0\times10^{-5}$  & $\phantom{+}0.25$ \\
Ca    & $7.7\pm0.8\times10^{-7}$  & $-0.42$ \\
Fe    & $3.7\pm0.8\times10^{-5}$  & $\phantom{+}0.12$ \\
\hline
\end{tabular} \\
\end{minipage}
\end{table}

\begin{table}
\centering
\begin{minipage}{85mm}
\caption{Accretion rates for selected elements.}
\label{tbl-7}
\centering
\renewcommand{\footnoterule}{\vspace*{-15pt}}
\begin{tabular}{ccc}
\hline
X & $\log\,\dot{M}$ (g\,s$^{-1}$)  &  [X/Si] \footnote{[X/Si]$\equiv\log{n({\rm X})/n({\rm Si})}-\log{n({\rm X})/n({\rm Si})}_\odot$} \\
\hline
C     & $<8.0$         & $-0.8$ \\
O     & $9.30\pm0.05$  & $\phantom{+}0.1$ \\
Mg    & $7.75\pm0.06$  & $-0.5$ \\
Si    & $8.25\pm0.07$  & ...    \\
Ca    & $6.39\pm0.04$  & $-0.9$ \\
Fe    & $8.45\pm0.09$  & $\phantom{+}0.0$ \\
\hline
\end{tabular} \\
\end{minipage}
\end{table}

\section{Discussion and conclusions}

We presented a model atmosphere analysis of the high-metallicity white dwarf GALEX~J1931+0117.
Based on a critical review of available atomic data, their uncertainties, and their impact on
the accuracy of abundance measurements, we revised the abundances presented by \citet{ven2010a}.
The variations are not significant with the exception of magnesium which is revised downward by
a factor of two. The abundance pattern shows that the abundance of magnesium, silicon, and iron
is near or above the solar abundance,
while the abundance of carbon, oxygen and calcium is below the solar abundance.
A line profile analysis performed using vertical abundance distributions obtained by solving the
steady-state diffusion equation shows that the accretion flow is rich in oxygen and silicon while
it is depleted in other elements. The calculations show that although the oxygen abundance is below solar in the
line forming region, it must be supplied in larger quantity because of its short
diffusion time-scale relative to other elements.

On the other hand, we show that introducing Stark shifts in line profile
modelling helped reduce the scatter in line velocity measurements to well within the expected
accuracy of the echelle spectra. Investigations of the effect of pressure shifts on
radial velocity measurements are relatively rare but their effect have been predicted \citep[e.g.,][]{ham1989,krs1992}.
A lack of radial velocity variations between two epochs that are 123 days apart
also rules out a close binary scenario
for the origin of the accreted material. We are left with the possibility that the infrared excess
belongs to a warm disc of debris material that accretes onto the white dwarf surface.

The iron abundance in GALEX~J1931+0117 is nearly solar \citep{asp2005} and it
is comparable to the iron abundance measured in much hotter white dwarfs
\citep{ven2006}. The resurgence of iron in cooler white dwarfs is also evidence that
an external reservoir reached the surface of these objects.

Simulated infrared spectral energy distributions for GALEXJ1931+0117 including
a spectral synthesis of the white dwarf added to a warm disc show that the
observed distribution \citep{ven2010a} can be matched by a 
warm disc (900 K) or a hot disc (1650 K) near sublimation temperature.
Necessarily, the inferred dust
temperature must be below sublimation temperatures \citep[see][]{lod2003} for the main constituents such as calcium (1659 K), or
silicon and magnesium (1354 K). 
Therefore, the temperature as well as the size and state (gas or solid) of the disc is
the result of an adjustment with the host white dwarf temperature and luminosity.
The corresponding disc sizes are 90$\Omega_{WD}$ for the hot disc, and, compensating for
its reduced emissivity per unit area, 2200$\Omega_{WD}$ for the warm
disc, where $\Omega_{WD}$ is the solid angle subtended by the white dwarf itself. 
The observed spectral energy distribution extends to the 2MASS measurements in the infrared
and measurements at longer wavelengths are required to further constrain the model.

A comparison with other warm DAZ white dwarfs supports these conjectures. The DAZ PG1015+161
\citep[$T_{\rm eff}=19300$ K, $\log{g}=7.9$, $\log{\rm Ca/H}=-5.9$,][]{koe2006} is similar
to GALEX1931+0117 ($T_{\rm eff}=20890$ K, $\log{g}=7.9$, $\log{\rm Ca/H}=-6.1$). Based
on JHK and Spitzer observations, \citet{jur2007} inferred the presence of a 
disc with an inner temperature ranging from 800 to 1000 K and emitting area
from 700 to 1200 $\Omega_{WD}$. In a similar case, \citet{far2009} inferred the presence of a disc with a blackbody temperature of 1500K around PG1457-086 ($T_{\rm eff}=20400$ K, $\log{g}=8.0$, $\log{\rm Ca/H}=-6.1$). The great diversity in disc emissivity is illustrated by the case of
HS0047+1903. Despite having a similar calcium abundance, no evidence of a disc is
found in JHK and Spitzer observations \citep{far2010b}. 

In conclusion, our analysis of the abundance pattern in GALEXJ1931+0117 
and the absence of a close companion, as well as a comparison with other, similar cases such as the DAZ PG1015+161
support the likely presence of a dusty disc around GALEXJ1931+0117. 

\section*{Acknowledgments}
S.V. and A.K. are supported by GA AV grant numbers IAA300030908 and IAA301630901, respectively, and by GA \v{C}R grant number P209/10/0967.
A.K. also acknowledges support from the Centre for Theoretical Astrophysics (LC06014). 
We thank M. Kilic for useful discussions and the referee D. Homeier for helpful comments.

\label{lastpage}


\begin{thebibliography}{99}
\bibitem[\protect\citeauthoryear{Artru et al.}{1981}]{art1981} Artru M.~C., Praderie F., Jamar C., Petrini D., 1981, A\&AS, 44, 171 
\bibitem[\protect\citeauthoryear{Asplund, Grevesse, \& Sauval}{2005}]{asp2005} Asplund M., Grevesse N., Sauval A.~J., 2005, ASPC, 336, 25 
\bibitem[\protect\citeauthoryear{Ben Nessib, Lakhdar, \& Sahal-Br{\'e}chot}{1996}]{ben1996} Ben Nessib N., Lakhdar Z.~B., Sahal-Br{\'e}chot S., 1996, PhyS, 54, 608 
\bibitem[\protect\citeauthoryear{Berengut, Dzuba, \& Flambaum}{2003}]{ber2003} Berengut J.~C., Dzuba V.~A., Flambaum V.~V., 2003, PhRvA, 68, 022502
\bibitem[\protect\citeauthoryear{Castelli}{2005}]{cas2005} Castelli F., 2005, Mem. S.A.It. Suppl., 8, 44 
\bibitem[\protect\citeauthoryear{Chayer, Fontaine, \& Wesemael}{1995}]{cha1995a} Chayer P., Fontaine G., Wesemael F., 1995, ApJS, 99, 189 
\bibitem[\protect\citeauthoryear{Chayer et al.}{1995}]{cha1995b} Chayer P., Vennes S., Pradhan A.~K., Thejll P., Beauchamp A., Fontaine G., Wesemael F., 1995, ApJ, 454, 429 
\bibitem[\protect\citeauthoryear{Debes}{2006}]{deb2006} Debes J.~H., 2006, ApJ, 652, 636 
\bibitem[\protect\citeauthoryear{Dimitrijevi{\'c} \& Sahal-Br{\'e}chot}{1992}]{dim1992} Dimitrijevi{\'c} M.~S., Sahal-Br{\'e}chot S., 1992, Bulletin Astronomique de Belgrade, 145, 83
\bibitem[\protect\citeauthoryear{Dimitrijevi{\'c} \& Sahal-Br{\'e}chot}{1995}]{dim1995} Dimitrijevi{\'c} M.~S., Sahal-Br{\'e}chot S., 1995, Bulletin Astronomique de Belgrade, 151, 101
\bibitem[\protect\citeauthoryear{Djeni{\v z}e, Sre{\'c}kovi{\'c}, \& Bukvi{\'c}}{2005}]{dje2005} Djeni{\v z}e S., Sre{\'c}kovi{\'c} A., Bukvi{\'c} S., 2005, JaJAP, 44, 1450 
\bibitem[\protect\citeauthoryear{Drullinger, Wineland, \& Bergquist}{1980}]{dru1980} Drullinger R.~E., Wineland D.~J., Bergquist J.~C., 1980, ApPhy, 22, 365 
\bibitem[\protect\citeauthoryear{Dupuis, Fontaine, \& Wesemael}{1993}]{dup1993} Dupuis J., Fontaine G., Wesemael F., 1993, ApJS, 87, 345
\bibitem[\protect\citeauthoryear{Dupuis et al.}{2009a}]{dup2009a} Dupuis J., Chayer P., H{\'e}nault-Brunet V., Vennes S., Kruk J.~W., 2009a, AIPC, 1135, 329 
\bibitem[\protect\citeauthoryear{Dupuis et al.}{2009b}]{dup2009b} Dupuis J., H{\'e}nault-Brunet V., Chayer P., Vennes S., Kruk J.~W., 2009b, JPhCS, 172, 012050 
\bibitem[\protect\citeauthoryear{Farihi et al.}{2010a}]{far2010a} Farihi J., Barstow M.~A., Redfield S., Dufour P., Hambly N.~C., 2010a, MNRAS, 404, 2123 
\bibitem[\protect\citeauthoryear{Farihi et al.}{2010b}]{far2010b} Farihi J., Jura M., Lee J.-E., Zuckerman B., 2010b, ApJ, 714, 1386 
\bibitem[\protect\citeauthoryear{Farihi, Jura, \& Zuckerman}{2009}]{far2009} Farihi J., Jura M., Zuckerman B., 2009, ApJ, 694, 805
\bibitem[\protect\citeauthoryear{Farihi, Zuckerman, \& Becklin}{2008}]{far2008} Farihi J., Zuckerman B., Becklin E.~E., 2008, ApJ, 674, 431 
\bibitem[\protect\citeauthoryear{Fontaine \& Michaud}{1979}]{fon1979} Fontaine G., Michaud G., 1979, ApJ, 231, 826
\bibitem[\protect\citeauthoryear{G{\"a}nsicke, Marsh, \& Southworth}{2007}]{gan2007} G{\"a}nsicke B.~T., Marsh T.~R., Southworth J., 2007, MNRAS, 380, L35
\bibitem[\protect\citeauthoryear{G{\"a}nsicke et al.}{2006}]{gan2006} G{\"a}nsicke B.~T., Marsh T.~R., Southworth J., Rebassa-Mansergas A., 2006, Sci, 314, 1908
\bibitem[\protect\citeauthoryear{Gonz{\'a}lez et al.}{2002}]{gon2002} Gonz{\'a}lez V.~R., Aparicio J.~A., del Val J.~A., Mar S., 2002, JPhB, 35, 3557
\bibitem[\protect\citeauthoryear{Graham et al.}{1990}]{gra1990} Graham J.~R., Matthews K., Neugebauer G., Soifer B.~T., 1990, ApJ, 357, 216 
\bibitem[Griem(1974)]{gri1974} Griem H.~R., 1974, Spectral line broadening by plasmas (Pure and Applied Physics Volume 39). Academic Press, Inc, New York
\bibitem[\protect\citeauthoryear{Hammond}{1989}]{ham1989} Hammond G.~L., 1989, LNP, 328, 346
\bibitem[\protect\citeauthoryear{Holberg, Barstow, \& Green}{1997}]{hol1997} Holberg J.~B., Barstow M.~A., Green E.~M., 1997, ApJ, 474, L127 
\bibitem[\protect\citeauthoryear{Hubeny \& Lanz}{1995}]{hub1995} Hubeny I., Lanz T., 1995, ApJ, 439, 875 
\bibitem[\protect\citeauthoryear{Jura, Farihi, \& Zuckerman}{2007}]{jur2007} Jura M., Farihi J., Zuckerman B., 2007, ApJ, 663, 1285 
\bibitem[\protect\citeauthoryear{Kawka et al.}{2008}]{kaw2008} Kawka A., Vennes S., Dupuis J., Chayer P., Lanz T., 2008, ApJ, 675, 1518 
\bibitem[\protect\citeauthoryear{Kilic et al.}{2006}]{kil2006} Kilic M., von Hippel T., Leggett S.~K., Winget D.~E., 2006, ApJ, 646, 474 
\bibitem[\protect\citeauthoryear{Koester}{2009}]{koe2009} Koester D., 2009, A\&A, 498, 517 
\bibitem[\protect\citeauthoryear{Koester et al.}{2005}]{koe2005} Koester D., Rollenhagen K., Napiwotzki R., Voss B., Christlieb N., Homeier D., Reimers D., 2005, A\&A, 432, 1025
\bibitem[\protect\citeauthoryear{Koester \& Wilken}{2006}]{koe2006} Koester D., Wilken D., 2006, A\&A, 453, 1051 
\bibitem[\protect\citeauthoryear{Kr{\v s}ljanin \& Dimitrijevi{\'c}}{1992}]{krs1992} Kr{\v s}ljanin V., Dimitrijevi{\'c} M.~S., 1992, LNP, 401, 371 
\bibitem[\protect\citeauthoryear{Kurucz \& Bell}{1995}]{kur1995} Kurucz R., Bell B., 1995, Atomic Line Data Kurucz CD-ROM No. 23. Smithsonian Astrophysical Observatory, Cambridge, MA
\bibitem[\protect\citeauthoryear{Lanz, Dimitrijevic, \& Artru}{1988}]{lan1988} Lanz T., Dimitrijevic M.~S., Artru M.-C., 1988, A\&A, 192, 249
\bibitem[\protect\citeauthoryear{Lanz \& Hubeny}{1995}]{lan1995} Lanz T., Hubeny I., 1995, ApJ, 439, 905 
\bibitem[\protect\citeauthoryear{Lesage}{2009}]{les2009} Lesage A., 2009, NewAR, 52, 471 
\bibitem[\protect\citeauthoryear{Lodders}{2003}]{lod2003} Lodders K., 2003, ApJ, 591, 1220
\bibitem[\protect\citeauthoryear{Matheron et al.}{2001}]{mat2001} Matheron P., Escarguel A., Redon R., Lesage A., Richou J., 2001, JQSRT, 69, 535 
\bibitem[\protect\citeauthoryear{Sre{\'c}kovi{\'c} \& Djeni{\v z}e}{1993}]{sre1993} Sre{\'c}kovi{\'c} A., Djeni{\v z}e S., 1993, Bulletin Astronomique de Belgrade, 148, 7 
\bibitem[\protect\citeauthoryear{Vennes et al.}{2006}]{ven2006} Vennes S., Chayer P., Dupuis J., Lanz T., 2006, ApJ, 652, 1554
\bibitem[\protect\citeauthoryear{Vennes, Kawka, \& N{\'e}meth}{2010a}]{ven2010a} Vennes S., Kawka A., N{\'e}meth P., 2010a, MNRAS, 404, L40 
\bibitem[\protect\citeauthoryear{Vennes, Kawka, \& N{\'e}meth}{2010b}]{ven2010b}
Vennes, S., Kawka, A., N{\'e}meth P., 2010b, in Schuh, S., Heber, U., 
Drechsel, H., eds, Planetary Systems beyond the Main Sequence, AIP, in press 
(arXiv:1012.2644)
\bibitem[\protect\citeauthoryear{Vennes et al.}{1988}]{ven1988} Vennes S., Pelletier C., Fontaine G., Wesemael F., 1988, ApJ, 331, 876 
\bibitem[\protect\citeauthoryear{Zuckerman et al.}{2003}]{zuc2003} Zuckerman B., Koester D., Reid I.~N., H{\"u}nsch M., 2003, ApJ, 596, 477

\end{thebibliography}
\end{document}